\documentclass[twocolumn,reprint,amsmath,amssymb,cha,aip]{revtex4-1}

\usepackage{docs}
\usepackage{epsfig}
\usepackage{graphicx}
\usepackage{bm}%
\usepackage[colorlinks=true,linkcolor=blue]{hyperref}%

\begin{document}

\title{Electronic structure of CaFe$_2$As$_2$}

\author{G. Adhikary}
\author{D. Biswas}
\author{N. Sahadev}
\author{R. Bindu}
\author{N. Kumar}
\author{S. K. Dhar}
\author{A. Thamizhavel}
\author{K. Maiti}
\altaffiliation{Correspondence to: kbmaiti@tifr.res.in}
\address {Department of Condensed Matter Physics and Materials' Science, Tata
Institute of Fundamental Research, Homi Bhabha Road, Colaba, Mumbai
- 400 005, INDIA.}

\begin{abstract}
We investigate the electronic structure of CaFe$_2$As$_2$ using high
resolution photoemission spectroscopy. Experimental results exhibit
three energy bands crossing the Fermi level making hole pockets
around the $\Gamma$-point. Temperature variation reveal a gradual
shift of an energy band away from the Fermi level with the decrease
in temperature in addition to the spin density wave (SDW) transition
induced Fermi surface reconstruction of the second energy band
across SDW transition temperature. The hole pocket in the former
case eventually disappears at lower temperatures while the hole
Fermi surface of the third energy band possessing finite $p$ orbital
character survives till the lowest temperature studied. These
results reveal signature of a complex charge redistribution among
various energy bands as a function of temperature that might be
associated to the exotic properties of this system.
\end{abstract}

\date{\today}


\maketitle

\section{Introduction}

Study of superconductivity has seen an explosive growth since the
discovery of high temperature superconductivity in cuprates in 1986
\cite{cuprate}. Discovery of superconductivity in Fe-based systems
\cite{Kamihara1, Kamihara2} led to a resurgence of interest in this
direction, where charge carrier doping in the parent compounds
having spin density wave (SDW) states to superconductivity via
suppression of long range magnetic order \cite{Iron-Pnictide}.
Interestingly, some of these Fe-based compounds also exhibit
pressure induced superconductivity \cite{pres} enhancing the
complexity of the problems in these materials.

Among the Fe-based superconductors, $A$Fe$_2$As$_2$ (A = Ca, Ba, Sr
and Eu) class of materials known as `122' compounds can be grown
easily with high quality and they have been studied extensively in
the recent past. These materials crystallize in the ThCr$_2$Si$_2$
type tetragonal structure at room temperature, (space group
$I4/mmm$). CaFe$_2$As$_2$ is one such compound exhibiting spin
density wave (SDW) transition due to the long range magnetic
ordering of the Fe moments at $T_{SDW}$ = 170 K along with a
concomitant structural transition to an orthorhombic phase. High
pressure \cite{pres}, substitution of Fe by Co, Ni\cite{Thamiz} and
other dopants induces superconductivity in CaFe$_2$As$_2$
(transition temperature, $T_c \sim$ 15 K). The SDW transition is
found to accompany a nesting of the Fermi surface
\cite{FSNesting,dessau} along with a transition from two dimensional
(2D) to three dimensional (3D) Fermi surface associated with the
structural transition \cite{3Dto2D-Kaminskii,Fink,Fink-EPL10}. Fe
3$d$ states play a major role in the electronic properties of these
systems, while the doped holes in cuprates possess dominant ligand
2$p$ orbital character \cite{cuprate}. Clearly, the physics of high
temperature superconductors is complex due to the significant
differences among different classes of materials. In this article,
we present our results on the electronic structure of CaFe$_2$As$_2$
obtained by high resolution photoemission spectroscopy and show the
signature of interesting gradual change in the electronic structure
with temperature that may be linked to the structural changes of
these materials. Such knowledge would be useful to design new such
materials with better specifications for potential applications.

\section{Experiment}

High quality single crystals of CaFe$_2$As$_2$ were grown using Sn
flux. The grown crystals are flat platelet like, which can be
cleaved easily and the cleaved surface looked mirror shiny. The
sample quality was verified by various characterization methods such
as $x$-ray diffraction (XRD) pattern and Laue pattern for
determining the crystal orientation, and energy dispersive analysis
of $x$-rays (EDAX) for composition analysis. The XRD and Laue
patterns were sharp and possess no spurious signal that confirmed
good crystallinity of the sample. EDAX results showed the sample to
be stoichiometric ensuring good sample quality. A sharp transition
to spin density wave state is observed at 170 K in both magnetic and
specific heat measurements.

Photoemission measurements were carried out using a Gammadata
Scienta analyzer, R4000 WAL and monochromatic photon sources, Al
$K\alpha$ ($h\nu$ = 1486.6 eV), He {\scriptsize I} ($h\nu$ = 21.2
eV) and He {\scriptsize II} ($h\nu$ = 40.8 eV) sources. The energy
resolution and angle resolution were set to 2 meV and 0.3$^o$
respectively for ultraviolet photoemission (UP) studies and the
energy resolution was fixed to 350 meV for $x$-ray photoemission
(XP) measurements. The temperature variation was carrier out using
an open cycle helium cryostat, LT-3M from Advanced Research Systems,
USA. The sample was cleaved {\it in situ} (base pressure
$<~3\times$~10$^{-11}$ Torr) at each temperature several times to
have a clean well ordered surface for the photoemission studies.
Reproducibility of the data in both cooling and heating cycle was
observed.

The energy band structure of CaFe$_2$As$_2$ was calculated using
full potential linearized augmented plane wave method within the
local density approximation (LDA) using Wien2k software
\cite{wien2k}. The energy convergence was achieved using 512
$k$-points within the first Brillouin zone. In order to calculate
the band structure for the paramagnetic phase, we used the lattice
parameters of the tetragonal phase; $a$ = 3.906\AA\ and $c$ = 12.124
\AA.

\section{Results}

An overview of the electronic structure involving the core levels
and the valence band have been captured by $x$-ray photoemission
spectroscopy with an energy resolution of 350 meV. Fe 2$p_{3/2}$
signal shown in Fig. 1(a) exhibits a highly asymmetric peak around
707 eV binding energy typical for metallic Fe-like behavior - the
asymmetry arises due to the screening of the core holes in the
photoemission final states and associated low energy excitations
\cite{acker,takahashi}. As 3$d$ spectra shown in Fig. 1(b) are also
asymmetric exhibiting two distinct features around 40.9 and 41.6 eV
binding energies due to the spin-orbit splitting. The feature at 44
eV binding energy is due to Ca 3$s$ core level photoemission. The
origin of the weak shoulder at about 40.4 eV binding energy can be
attributed to the surface effects \cite{surf}. Fe 2$p$ and As 3$d$
spectra at 10 K and 300 K are almost identical indicating unchanged
Madelung potential and/or effective valency of Fe and As in the
whole temperature range despite the structural transition from the
tetragonal to orthorhombic phase at 170 K.

\begin{figure}
 \vspace{-2ex}
\includegraphics[scale=0.45]{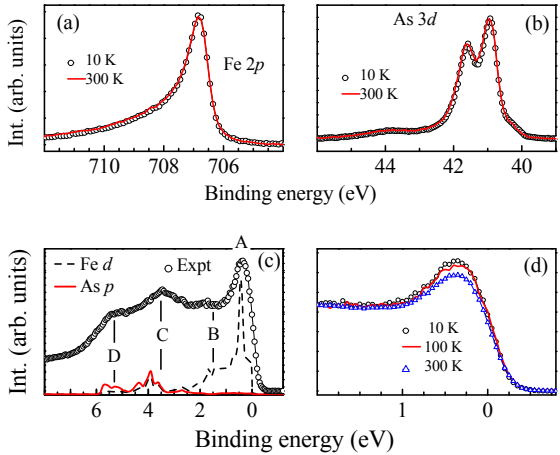}
 \vspace{-42ex}
 \caption{(a) Fe 2$p$ and (b) As 3$d$ core level spectra at 10 K and 300 K.
(c) Valence band spectrum at 300 K (symbols). The dashed line
represents Fe 3$d$ and the solid line represents As 4$p$ partial
density of states. (d) Evolution of the valence band spectra near
$\epsilon_F$ with temperatures.}
 \vspace{-2ex}
\end{figure}

The valence band spectrum collected using Al $K\alpha$ photon source
at 300 K is shown in Fig. 1(c). It exhibits four distinct features
denoted by A, B, C and D. A remarkably good representation of the
experimental spectrum is given by the energy band structure of the
tetragonal CaFe$_2$As$_2$ calculated by employing full potential
linearized augmented plane wave method within the local density
approximation (LDA) using Wien2k software \cite{wien2k}. The orbital
polarization of the energy bands are obtained by projecting the
eigen states onto the constituting electronic states, namely the Fe
3$d$ and As 4$p$ states in the present case. The calculated partial
density of states (PDOS) shown by lines in the figure reproduce well
the experimental spectrum. The feature A exhibits dominant Fe 3$d$
character and the As 4$p$ states contribute at higher binding
energies. The hybridization between As 4$p$ and Fe 3$d$ states is
found to be strong. The temperature evolution of the valence band is
shown in Fig. 1(d). A normalization by the spectral intensities in
the energy range beyond 1 eV binding energy exhibits a gradual
enhancement of the feature A intensity with the decrease in
temperature.

\begin{figure}
 \vspace{-2ex}
\includegraphics[scale=0.45]{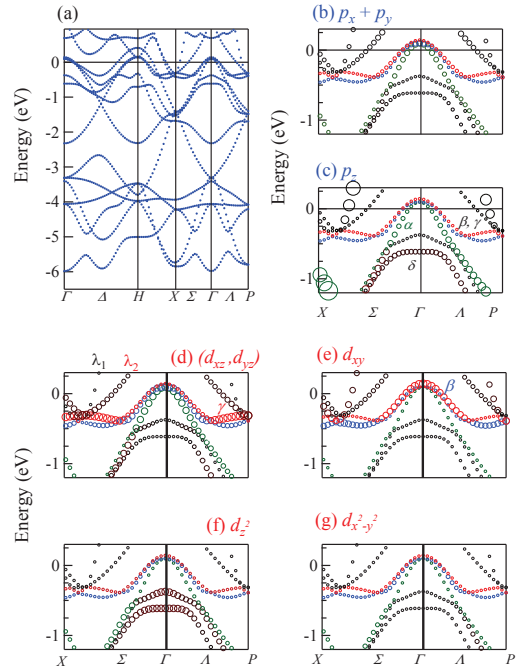}
 \vspace{-26ex}
 \caption{(a) Energy band dispersions calculated using FLAPW method.
The orbital contributions of the bands near the Fermi level are
shown by the size of the symbols for (b) As $p_x$+$p_y$, (c) As
$p_z$, (d) Fe (3$d_{xz}$,3$d_{yz}$) (e) Fe 3$d_{xy}$, (f) Fe
3$d_{z^2}$ and (g) Fe 3$d_{x^2-y^2}$ states. In order to have
clarity on As $p$ contributions, the corresponding symbol sizes are
renormalized.}
 \vspace{-2ex}
\end{figure}

The details of the calculated energy bands and their orbital
characters are shown in Fig. 2. The size of the symbols represents
the orbital contributions. The energy bands near the Fermi level,
$\epsilon_F$ exhibit significant dispersion ($\geq$ 0.5 eV). Three
bands denoted by $\alpha$, $\beta$ and $\gamma$ cross $\epsilon_F$
around the $\Gamma$ point forming three hole-pockets. $\lambda_1$
and $\lambda_2$ bands form the electron pockets around the
$X$-point. Fig. 2(b) and 2(c) show finite As 4$p$ contributions in
these energy bands arising due to the Fe 3$d$ - As 4$p$
hybridizations. In Figs. 2(d) - 2(g), we show the ($d_{xz},d_{yz}$),
$d_{xy}$, $d_{z^2}$ and $d_{x^2-y^2}$ contributions, respectively.
$\alpha$ and $\gamma$ bands exhibit large ($d_{xz},d_{yz}$)
symmetry, while the $\beta$ band possesses primarily $d_{xy}$
symmetry. $d_{z^2}$ states appear as $\delta$ bands shown in Fig.
2(f). The Fermi surfaces corresponding to $\lambda_2$ and $\gamma$
bands possessing similar symmetry are well nested.

\begin{figure}
 \vspace{-2ex}
\includegraphics[scale=0.45]{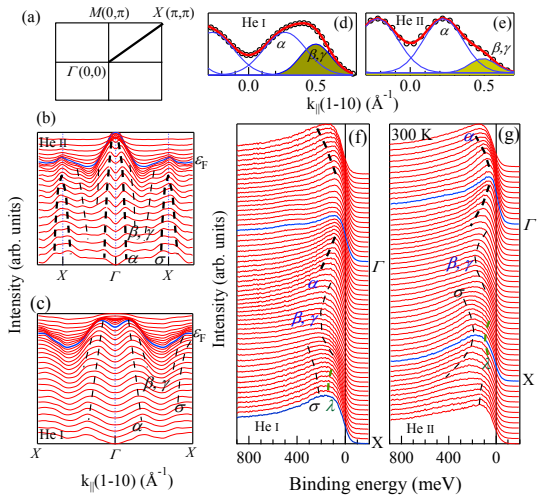}
 \vspace{-40ex}
 \caption{(a) A typical Brillouin zone of CaFe$_2$As$_2$. MDCs
at 300 K employing (b) He {\scriptsize II} and (c) He {\scriptsize
I} photon energies. A typical fit of the MDC at 140 meV to find peak
positions in the (d) He {\scriptsize I} and (e) He {\scriptsize II}
photon energies. EDCs at 300 K using (f) He {\scriptsize I} and (g)
He {\scriptsize II} photon energies.}
 \vspace{-2ex}
\end{figure}

The momentum distribution curves (MDCs) along $\Gamma X$ direction
[see the Brillouin zone in Fig. 3(a)] obtained at 300 K exhibit
significant differences when probed with He {\scriptsize II} and He
{\scriptsize I} photon energies as shown in Figs. 3(b) and 3(c),
respectively. Crystal structure of CaFe$_2$As$_2$ is tetragonal at
300 K and it exhibits an essentially two-dimensional Fermi surface
\cite{3Dto2D-Kaminskii}. Thus, different photon energies differing
in the $k_z$ values ($k_z~\sim~$9.5$\pi/c$ for He {\scriptsize I}
and $\sim~$12.5$\pi/c$ for He {\scriptsize II}) will have a weak
influence on the energy position of the spectral features. In any
case, the change in the relative intensity of the spectral features
with the change in photon energy arises primarily due to the matrix
element effect associated with the photo-excitation process that
reflects the orbital character of the features \cite{yeh}. The
simulations of each of the MDCs require at least two peaks
representing $\alpha$-band and ($\beta, \gamma$)-bands as shown in
Figs. 3(d) and 3(e) for the MDCs at 140 meV binding energy. The
large intensity of the $\alpha$-band in the He {\scriptsize
II}-spectra and its reduction with respect to the intensity of
($\beta, \gamma$)-bands in the He {\scriptsize I} spectra suggests
larger As 4$p$-contribution in ($\beta, \gamma$)-bands relative to
that in the $\alpha$-band. This behavior is consistent with the
observation in other compounds in the same class of materials
\cite{EuFe2As2}. The influence of the matrix element is also evident
in the energy distribution curves (EDCs) shown in Figs. 3(f) and
3(g).

The $\alpha$, $\beta$ and $\gamma$ bands form hole-pockets around
the $\Gamma$ point at 300 K. The $\delta$ bands possessing
$d_{z^2}$-symmetry appear around 400 meV below $\epsilon_F$. At
$X$-point, signature of the two bands denoted by $\sigma$ and
$\lambda$ are observed in the figure. The $\lambda$ band forms an
electron pocket around $X$-point. The bandwidth of the
experimentally observed energy bands is significantly smaller than
the calculated ones manifesting the signature of electron
correlation induced effects in the electronic structure.

\begin{figure}
 \vspace{-2ex}
\includegraphics[scale=0.45]{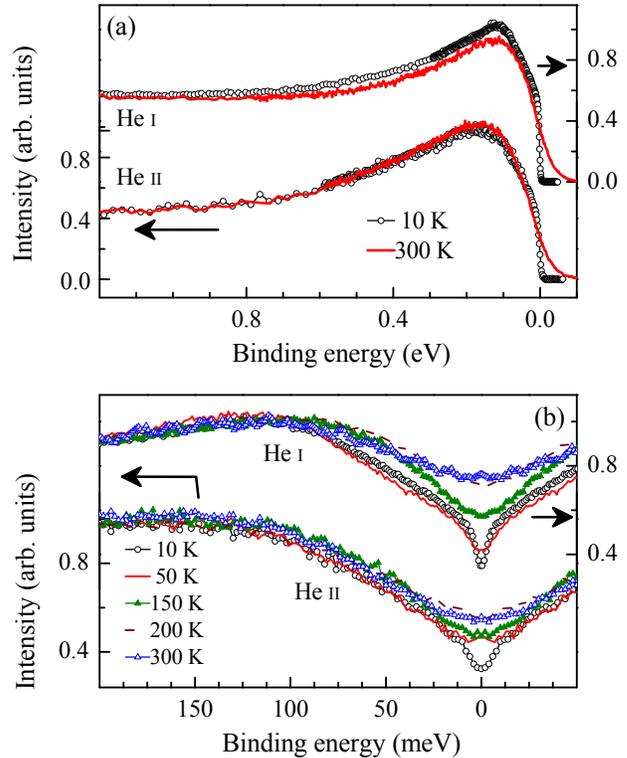}
 \vspace{-12ex}
 \caption{(a) High resolution photoemission spectra at
10 K and 300 K using He {\scriptsize I} and He {\scriptsize II}
photon energies. (b) The symmetrized spectral density of states
(SDOS) of the He {\scriptsize I} and He {\scriptsize II} spectra.}
 \vspace{-2ex}
\end{figure}

In order to investigate the temperature evolution of the electronic
states at $\epsilon_F$ critically, we employed high energy
resolution and the angle integrated mode for better signal to noise
ratio (acceptance angle $\pm$15$^o$) setting the $k$-vector along
the $\Gamma X$ direction. In such a case, the intensity at
$\epsilon_F$ will be contributed by the energy bands crossing the
Fermi level. The photoemission spectra obtained by He {\scriptsize
I} and He {\scriptsize II} photon energies at 10 K and 300 K are
shown in Fig. 4(a). All the spectra are normalized by the intensity
beyond 1 eV binding energy. The spectral changes are most profound
in the He {\scriptsize I} spectra while the He {\scriptsize II}
spectra exhibit modification essentially at $\epsilon_F$. This is
attributed to the higher degree of thermal sensitivity of the
electronic states with $p$ orbital character relative to that of the
$d$ character.

\begin{figure}
 \vspace{-2ex}
\includegraphics[scale=0.45]{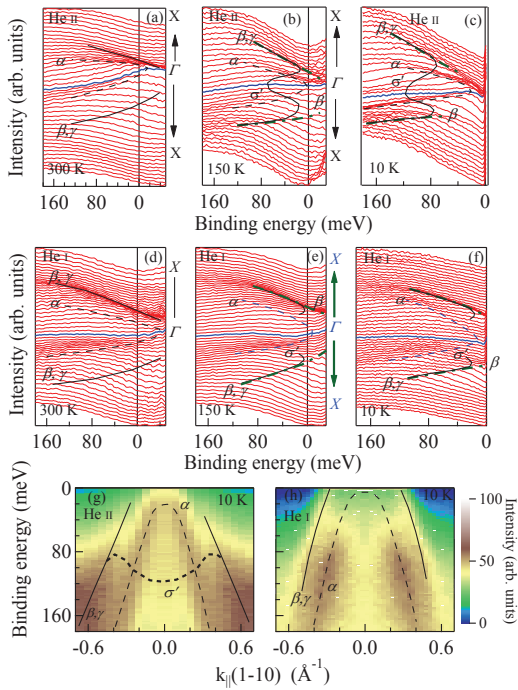}
 \vspace{-8ex}
 \caption{He {\scriptsize II} EDCs along $\Gamma X$ at (a) 300 K, (b) 150 K
and (c) 10 K. He {\scriptsize I} EDCs along $\Gamma X$ at (d) 300 K,
(e) 150 K and (f) 10 K. Color plot of the same bands at 10 K using
(g) He {\scriptsize II} and (h) He {\scriptsize I} energies.}
 \vspace{-2ex}
\end{figure}

The spectral intensity at $\epsilon_F$ can be estimated quite
accurately by symmetrization of the experimental spectra
($I(\epsilon) = I_{expt}(\epsilon - \epsilon_F) +
I_{expt}(\epsilon_F - \epsilon)$; $I_{expt}$ is the experimental
spectrum). The symmetrized spectra at different temperatures, shown
in Fig. 4(b), exhibit a jump in the intensity at $\epsilon_F$ across
the SDW transition temperature in both He {\scriptsize I} and He
{\scriptsize II} spectra as expected. Further lowering of
temperature leads to an unusually sharp dip below 100 K although no
phase/structural transitions have been reported in this temperature
range.

The details of the changes can be found in the angle resolved data.
The energy bands obtained by He {\scriptsize I} and He {\scriptsize
II} excitations are shown in Fig. 5 along $\Gamma X$ direction at
300 K, 150 K and 10 K. The spectra exhibit an interesting evolution
across the SDW transition at 170 K. The signature of the hole pocket
corresponding to the $\alpha$ band distinctly seen in the 300 K He
{\scriptsize II} spectra vanishes at 150 K with the top of the band
appearing below $\epsilon_F$. In addition, the intensity around 80
meV grows significantly in the 150 K data relative to the 300 K data
suggesting SDW phase transition induced band-folding of the $\gamma$
and $\lambda$ bands consistent with the literature \cite{Lifschitz}.
Such folded band denoted by $\sigma^\prime$ in the figure appears
due to the nesting of the Fermi surfaces corresponding to the
$\gamma$ and $\lambda$ bands in the SDW phase. The signature of the
$\sigma^\prime$ band is also observed in the He {\scriptsize I}
spectra shown in Fig. 5(e). However, the hole pocket corresponding
to the $\alpha$-band survives in the He {\scriptsize I} spectra
($k_z \approx 9.5\pi/c$), which is different from the He
{\scriptsize II} spectra ($k_z \approx 12.5\pi/c$). This reflects
the $k_z$-dependence of the $\alpha$-band Fermi surface undergoing
transition from the effective two dimensional nature of the Fermi
surface to a three dimensional one with the $\alpha$-band hole
pocket centering around $k_z~\sim~2(2n+1)\pi/c$ and its absence
around $4n\pi/c$ in the $k$-plane containing $k_z$ axis
\cite{3Dto2D-Kaminskii,dessau}.

Further decrease in temperature reveals an unusual behavior; the
$\alpha$ band shifts further below in energy. The gradual energy
shift is most evident in the spectral density of states (SDOS)
obtained from He {\scriptsize II} spectra shown in Fig. 5(c), where
the Fe 3$d$ states contribution is significantly enhanced in the
photoemission signal due to the matrix element effect. The top of
the $\alpha$-band moves below the Fermi level at 10 K even in the He
{\scriptsize I} spectra shown in Figs. 5(f) and 5(h) suggesting
vanishing of the corresponding Fermi surface even at about
9.5$\pi/c$, which is close to the middle of the $\alpha$-band hole
pocket in reciprocal plane containing $k_z$-axis. The $\beta$-band
still crosses $\epsilon_F$, which is observed most prominently in
the He {\scriptsize I} spectra indicating its large $p$ character.

\section{Discussion}

Treating electron correlation induced effects in the strongly
correlated electron systems has been a major challenge in the
contemporary condensed matter physics. It is observed that the
spectral functions obtained by the dynamical mean field theory often
provide a good description of the experimental scenario
\cite{RMP1,RMP2}. Within this description, the spectral function of
a correlated system in the intermediate coupling regime consists of
three features; upper (unoccupied) and lower (occupied) Hubbard
bands representing the correlation induced localized states and a
central band at the Fermi level, termed as the coherent feature
representing the itinerant carriers often captured well by the
\emph{ab initio} band structure calculations \cite{csvo}. The
decrease in temperature leads to an enhancement of the coherent
feature intensity at the cost of the Hubbard band intensities
\cite{RMP1,ruth}. The temperature induced changes in the valence
band shown in Fig. 1 exhibit quite similar scenario, where the
intensity at the Fermi level, well reproduced in the band structure
calculations, increases in intensity relative to the intensities at
higher binding energies with the decrease in temperature. Effect of
the electron correlation is also manifested by the narrowing of the
energy bands shown in Fig. 3.

We calculated the effective mass, $m^*/m_e$ ($m_e$ = mass of an
electron) corresponding to $\alpha$ and $\beta,\gamma$ bands, and
found it to be 1.4 and 3.2, respectively at 300 K suggesting
moderate mass enhancement due to electron correlation induced
effects supporting the above view. The energy bands close to the
Fermi level are the antibonding bands arising due to the
hybridization between Fe 3$d$ and As 4$p$ states. While these eigen
states possess large Fe 3$d$ character, As 4$p$ contributions appear
to be significant, which is observed in the band structure results
shown in Fig. 2 and comparison of the experimental He {\scriptsize
I} \& He {\scriptsize II} spectra shown in Fig. 3. Since the As
layers are above and below the Fe-layers, the hybridization of As
$p$ states with the Fe $d_{xz}, d_{yz}$ states will be strong and
plays an important role in the SDW phase transition. This is
manifested via the nesting of the Fermi surface corresponding to the
$\gamma$ and $\lambda$ bands possessing ($d_{xz}, d_{yz}$) symmetry.
Signature of the Fermi surface nesting observed in the present
results is consistent with the earlier studies. The importance of
the coupling of the electronic states with the lattice degrees of
freedom has been found in this class of compounds via inverse
isotope effect \cite{Shirage}, spin-phonon coupling
\cite{CaFeAsSood,FeAsSood}, phonon softening \cite{CaFeAsOFSood},
sensitivity of the SDW phase to the anomalous phonon dispersion
\cite{phonon-mittal} etc. The photoemission and band structure
results provide a direct evidence of the importance of the
hybridization and the electron correlation effects in the electronic
properties of these systems.

At 150 K, which is below the SDW transition temperature, the top of
the $\alpha$ band appears below $\epsilon_F$ in the He {\scriptsize
II} spectra ($k_z~\sim~12.5\pi/c$) while it is still crossing
$\epsilon_F$ in the He {\scriptsize I} spectra
($k_z~\sim~9.5\pi/c$). This suggests a $k_z$-dependence of the
$\alpha$-band Fermi surface, which is a signature of the three
dimensionality of the Fermi surface consistent with the earlier
findings \cite{3Dto2D-Kaminskii}. Such $k_z$ dependence can be
attributed to the structural transition from tetragonal to
orthorhombic phase occurring at the same SDW transition temperature.
While energy bands at 300 K are narrower than their band structure
results due to electron correlation induced effects, the spectra at
10 K exhibit further narrowing of the dispersions. We calculated the
effective mass, $m^*/m_e$ corresponding to $\alpha$ and $\gamma$
bands and found it to increase to 2.6 and 4.3 at 10 K, respectively.
This suggests that the electrons acquire more local character at
lower temperatures.

\begin{figure}
 \vspace{-2ex}
\includegraphics[scale=0.45]{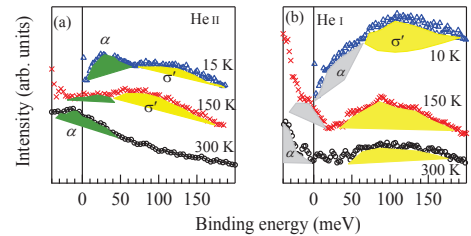}
 \vspace{-60ex}
 \caption{Temperature evolution of the (a) He {\scriptsize II} and
(b) He {\scriptsize I} spectra at the $\Gamma$ point.}
 \vspace{-2ex}
\end{figure}

Interestingly, the $\alpha$ band shifts below $\epsilon_F$ at 10 K
even in the He {\scriptsize I} spectra as shown in Fig. 5. This is
unusual as the He {\scriptsize I} energy corresponds to the $k_z$
value close to the middle of the $\alpha$-band hole pocket. In order
to exemplify this point, we compare the EDCs at $\Gamma$ point at
different temperatures obtained by He {\scriptsize I} and He
{\scriptsize II} photon energies in Fig. 6 exhibiting distinct
signature of a gradual shift of the $\alpha$ band with the decrease
in temperature. Sharp dip observed in the high resolution spectra
may be attributed to the shift of the $\alpha$ band completely below
$\epsilon_F$ which means that the $\alpha$-band Fermi surface
disappears around 10 K.

Two conclusions can be drawn from the above results. (i) Although
the structural transition from tetragonal to orthorhombic phase
occurs at the onset of SDW phase transition, the transition is slow
with the coexistence of two structural phases in the intermediate
temperature range as observed in other systems such as manganites
\cite{bindu}. (ii) The second observation is the signature of the
vanishing of the Fermi surface corresponding to the $\alpha$ band
and its consequences on the energy band, $\beta$ that crosses the
Fermi level in order to preserve charge count.

It is to note here that Lifshitz transition, a transition from a
hole-type Fermi surface to electron-type Fermi surface or vice versa
due to subtle change in charge carrier concentration has been found
to be important in cuprates \cite{thomas} as well as in electron
doped Ba(Fe$_{1-x}$Co$_x$)$_2$As$_2$ \cite{Lifschitz}. Evidently,
such Fermi surface reconstruction seem to be important in
CaFe$_2$As$_2$ too suggesting a generic nature of such an effect in
the unconventional superconductors. It is to be noted here that a
homologous system, SrFe$_2$As$_2$ \cite{srfe2as2} exhibits large
volume fraction of superconducting region even at ambient pressure
if the prepared sample possess significant structural strain. In
CaFe$_2$As$_2$, it is observed\cite{pres} that a quasi-hydrostatic
pressure leads to superconductivity suggesting the importance of
structural strain for such exotic ground state.

\section{Conclusions}

In summary, we have studied the evolution of the electronic
structure of CaFe$_2$As$_2$, a parent compound of the Fe-based
superconductors employing high resolution photoemission
spectroscopy. We discover a sharp dip at the Fermi level appearing
much below the SDW transition temperature in addition to the dip
associated with the SDW transition. Angle resolved photoemission
results exhibit signature of a gradual shift of the $\alpha$ band
with the decrease in temperature - the hole pocket corresponding to
the $\alpha$ band vanishes near 10 K indicating renormalization of
the $\beta$ band hole pocket to preserve charge count. These
results, thus, reveal signature of interesting Fermi surface
reconstructions in these complex systems that might be responsible
for exotic properties of these materials.

\section{Acknowledgements}

The authors, K. M. and N. S. acknowledge financial support from the
Department of Science and Technology, Government of India under the
`Swarnajayanti Fellowship programme'. K. M. acknowledges the
Department of Atomic Energy, Government of India for financial
support.

\end{document}